\documentclass[cits]{PoS}

\usepackage{wrapfig}

\newcommand{\CSSM}{Special Research Centre for the Subatomic Structure
  of Matter (CSSM),\\School of Chemistry and Physics, University of
  Adelaide, South Australia 5005, Australia} 

\newcommand{\CoEPP}{Special Research Centre for the Subatomic Structure
  of Matter (CSSM) and\\ARC Centre of Excellence for Particle Physics at
  the Terascale (CoEPP),\\School of Chemistry and Physics, University
  of Adelaide, South Australia 5005, Australia} 

\newcommand{\NCI}{National Computational Infrastructure
  (NCI),\\Australian National University, Australian Capital Territory
  0200, Australia}

\title{On the Structure of the Lambda 1405}

\ShortTitle{On the Structure of the Lambda 1405}

\author{Jonathan M. M. Hall, Waseem Kamleh\\
        \CSSM }

\author{\speaker{Derek B. Leinweber}%
         \thanks{We thank PACS-CS Collaboration for making
their $2+1$ flavor configurations available and acknowledge the
ongoing support of the ILDG.
This research was undertaken with the assistance of resources at the
NCI National Facility in Canberra, Australia, and the iVEC facilities
at Murdoch University (iVEC@Murdoch) and the University of Western
Australia (iVEC@UWA). These resources are provided through the
National Computational Merit Allocation Scheme and the University of
Adelaide Partner Share supported by the Australian Government.  We
also acknowledge eResearch SA for their support of our supercomputers.
This research is supported by the Australian Research Council through
the ARC Centre of Excellence for Particle Physics at the Terascale,
and through Grants DP120104627 (DBL), DP140103067 (DBL and
RDY), FT120100821 (RDY) and FL0992247 (AWT).}\\
        \CSSM \\
        E-mail: \email{derek.leinweber@adelaide.edu.au}}

\author{Benjamin J. Menadue\\
        \CSSM \\
        \NCI}

\author{Benjamin J. Owen\\
        \CSSM }

\author{Anthony W. Thomas, Ross D. Young\\
        \CoEPP }

\abstract{ For almost 50 years the structure of the \boldmath{$\Lambda(1405)$}
  resonance has been a mystery.  Recently, a new lattice QCD
  simulation showing that its strange magnetic form factor vanishes,
  together with a comprehensive Hamiltonian analysis of the lattice
  QCD energy levels, has unambiguously established that the structure
  is dominated by a bound anti-kaon--nucleon component
  \cite{Hall:2014lambda}.  Here we present supplementary information
  for Ref.~\cite{Hall:2014lambda} including a presentation of the
  relevant Hamiltonian effective field theory and an illustration of
  the volume dependence of the results and their connection to the
  infinite volume limit of Nature.}

\FullConference{The 32nd International Symposium on Lattice Field Theory\\
		 23-28 June, 2014\\
		 Columbia University New York, NY}

\begin{document}

\section{Introduction}

For almost 50 years the structure of the $\Lambda(1405)$ resonance has
been a mystery.  Even though it contains a relatively massive strange
quark and has odd parity, {\em both of which should increase its
  mass}, it is, in fact, lighter than any other excited spin-1/2
baryon.  Identifying the explanation for this observation has
challenged theorists since its discovery in the
1960s~\cite{Engler:1965zz}.
While the quantum numbers of the $\Lambda(1405)$ can be described by
three quarks, $(uds)$, its totally unexpected position in the spectrum
has rendered its structure quite mysterious.

Recently, the very first Lattice QCD calculation of the
electromagnetic form factors of the $\Lambda(1405)$ were presented
\cite{Hall:2014lambda}.  This calculation reveals the vanishing of the
strange quark contribution to the magnetic form factor of the
$\Lambda(1405)$ in the regime where the masses of the up and down
quarks approach their physical values.  

This result is very naturally
explained if the state becomes a molecular $\overline{K}N$ bound state
in that limit.  
In this case the strange quark is confined within a spin-0 kaon and
has no preferred spin orientation. Because the anti-kaon has zero
orbital angular momentum in order to conserve parity, the strange
quark {\em cannot} contribute to the magnetic form factor of the
$\Lambda(1405)$.
On the other hand, if the $\Lambda(1405)$ were a $\pi \Sigma$ state or
an elementary three-quark state the strange quark must make a sizable
contribution to the magnetic form factor.  Only if the $\overline{K}N$
component in the structure of the $\Lambda(1405)$ is dominant would
one expect to find a vanishing strange-quark magnetic form factor.

Upon combining the observation of a vanishing strange quark magnetic
form factor with a Hamiltonian effective-field-theory analysis of the
structure of the state as a function of its light quark mass, which
shows $\overline{K}N$ dominance and a rapidly decreasing wave function
renormalisation constant in the same limit, one concludes that this
long-standing problem has been unambiguously resolved
\cite{Hall:2014lambda}.

In this presentation we provide additional information in support of
Ref.~\cite{Hall:2014lambda} including a presentation of the relevant
Hamiltonian effective field theory and an illustration of the volume
dependence of the results and their connection to the infinite volume
limit of Nature.

\section{Hamiltonian Model}

In this section we present the construction and analysis of the
finite-volume Hamiltonian effective field theory.  We follow a
procedure similar to that for the $\Delta \to N \pi$ analysis
\cite{Hall:2013qba}.  However, in the present case multiple
meson-baryon channels are considered.  In constructing the
Hamiltonian, the four octet meson-baryon interaction channels of the
$\Lambda(1405)$ are included: $\pi\Sigma$, $\overline{K}N$, $K\Xi$,
$\eta\Lambda$.

We begin by writing the Hamiltonian $H$ as the sum of free and
interacting Hamiltonians, $H = H_0 + H_I$.  The rows and columns of
$H$ represent the magnitudes of the three-momenta available to the
meson-baryon intermediate states dressing the bare $\Lambda(1405)$
state.  As we work with total momentum zero, the meson and the baryon
will each carry the same magnitude of momentum in a back-to-back
orientation.
In a finite periodic volume, momentum is quantised.  Working on a
cubic volume of extent $L$ on each side, it is convenient to define
the momentum magnitudes
\begin{equation}
k_n = \sqrt{n_x^2 + n_y^2 + n_z^2} \, \frac{2\pi}{L} \, ,
\end{equation}
with $n_i = 0, 1, 2, \ldots$ and integer $n = n_x^2 + n_y^2 + n_z^2$.
As there are permutations of the $n_i$ that give rise to the same
momentum magnitude, we also introduce the integer $C_3(n)$ as a
combinatorial factor equal to the number of unique permutations of
$\pm n_x$, $\pm n_y$ and $\pm n_z$.  For example, the lowest lying
nontrivial momentum available on the lattice, where one direction has
the magnitude of $2 \pi / L$, has $C_3(1) = 6$.  The result recognises
three positions for the non-trivial momentum and a factor of two
associated with whether the meson or the baryon carries the positive
momentum.

The non-interacting Hamiltonian $H_0$ has diagonal entries
corresponding to the relativistic non-interacting meson-baryon
energies available on the finite periodic volume at total
three-momentum zero.  It also includes a single-particle state with
bare mass, $m_0 + \alpha_0\, m_\pi^2$.
The parameters $m_0$ and $\alpha_0$ are to be constrained by the
lattice QCD results.
Denoting each meson-baryon energy by $\omega_{MB}(k_n) = \omega_M(k_n)
+ \omega_B(k_n)$, with $\omega_A(k_n) \equiv \sqrt{k_n^2 + m_A^2}$,
the non-interacting Hamiltonian takes the form
\begin{equation}
\small
\label{eq:H0}
 H_0  = \left (
\begin{array}{cccc}
m_0 + \alpha_0\, m_\pi^2 & 0 & 0 & \cdots\\
0 &
\begin{array}{ccc}
\omega_{\pi\Sigma}(k_0)&&\\ &\ddots&\\&&\omega_{\eta\Lambda}(k_0)
\end{array}
  &0 &\cdots\\
0 & 0 &
\begin{array}{ccc}
\omega_{\pi\Sigma}(k_1)&&\\ &\ddots&\\&&\omega_{\eta\Lambda}(k_1)
\end{array}& \cdots\\
\vdots & \vdots & \vdots  & \ddots
\end{array}
\right )
\, .
\end{equation}
In the present model the interaction entries describe the coupling of
the single-particle state to the two-particle meson-baryon states.
\begin{equation}
\small
\label{eq:HI}
 H_I  = \left (
\begin{array}{ccccccc}
0& g_{\pi\Sigma}(k_0) &\cdots& g_{\eta\Lambda}(k_0) &g_{\pi\Sigma}(k_1)
&\cdots& g_{\eta\Lambda}(k_1)\cdots  \\
g_{\pi\Sigma}(k_0) & 0 & \cdots &  &&& \\
\vdots  & \vdots & 0 & &&&\\
g_{\eta\Lambda}(k_0) & & & \ddots&&&\\
g_{\pi\Sigma}(k_1) & & & &&&\\
\vdots & & & &&&\\
g_{\eta\Lambda}(k_1) & & & &&&\\
\vdots & & & &&&
\end{array}
\right )
\, .
\end{equation}
Each entry represents the $S$-wave interaction energy of the
$\Lambda(1405)$ with one of the four channels at a certain value for
$k_n$.
The form of the interaction is derived from effective field theory,
and includes the relevant finite-volume factors \cite{Hall:2013qba}
\begin{equation}
g_{MB}(k_n) = \left(\frac{\kappa_{MB}}{16\pi^2 f_\pi^2}\frac{C_3(n)}{4\pi}
\left(\frac{2\pi}{L}\right)^{3} \omega_M(k_n)\, u^2(k_n)
\right)^{1/2}. 
\label{eq:gMB}
\end{equation}
$f_\pi = 92.4$ MeV represents the pion decay constant and $u^2(k_n)$
is a regulator function.
For the purposes of this model, $u(k_n)$ takes the form of a dipole
regulator, with a regularization scale of $\Lambda = 0.8$ GeV.
It has been shown in previous investigations that the regulator
dependence is small in the extraction of resonance parameters near the
physical pion mass
\cite{Leinweber:2003dg}.
The coupling $\kappa_{MB}$ is related to the $SU(3)$-flavour singlet
couplings of the octet mesons and baryons via \cite{Veit:1984an,Veit:1984jr} 
\begin{equation}
\kappa_{\pi\Sigma} = 3 \xi_0,\qquad \kappa_{\bar{K}N} = 2 \xi_0,\qquad
\kappa_{K\Xi} = 2 \xi_0,\qquad \kappa_{\eta\Lambda} = \xi_0,  
\end{equation}
with $\xi_0 = 0.75$, chosen in order to ensure that the 
$\pi\Sigma$ decay width of the $\Lambda(1405)$ takes the physical
value of $50\pm 2$ MeV \cite{Beringer:1900zz} at the physical pion
mass, in the infinite-volume limit.

The eigenvalue equation corresponding to the Hamiltonian model
presented here takes the following simple form which is consistent
with chiral effective field theory
\begin{equation}
\lambda = m_0 + \alpha_0\, m_\pi^2 - \sum_{M,B}\sum_{n=0}^\infty 
\frac{g_{MB}^2(k_n)}{\omega_{MB}(k_n)-\lambda} \, .
\label{eq:eval1}
\end{equation}
$M$ and $B$ denote the intermediate meson-baryon pairs, with
coupling $g_{MB}(k_n)$ provided in Eq.~(\ref{eq:gMB}).
$\lambda$ is the energy eigenvalue of the $S$-wave interaction, which
occurs on both sides of the equation.  As $\lambda$ is finite, the
pole in the denominator of the right-hand side is never accessed.  A
nontrivial mixing of states occurs and the bare mass $m_0 + \alpha_0\,
m_\pi^2$ encounters self-energy corrections that lead
to avoided level-crossings in the finite-volume energy eigenstates.

\begin{figure}[t]
\includegraphics[height=0.48\columnwidth,angle=90]{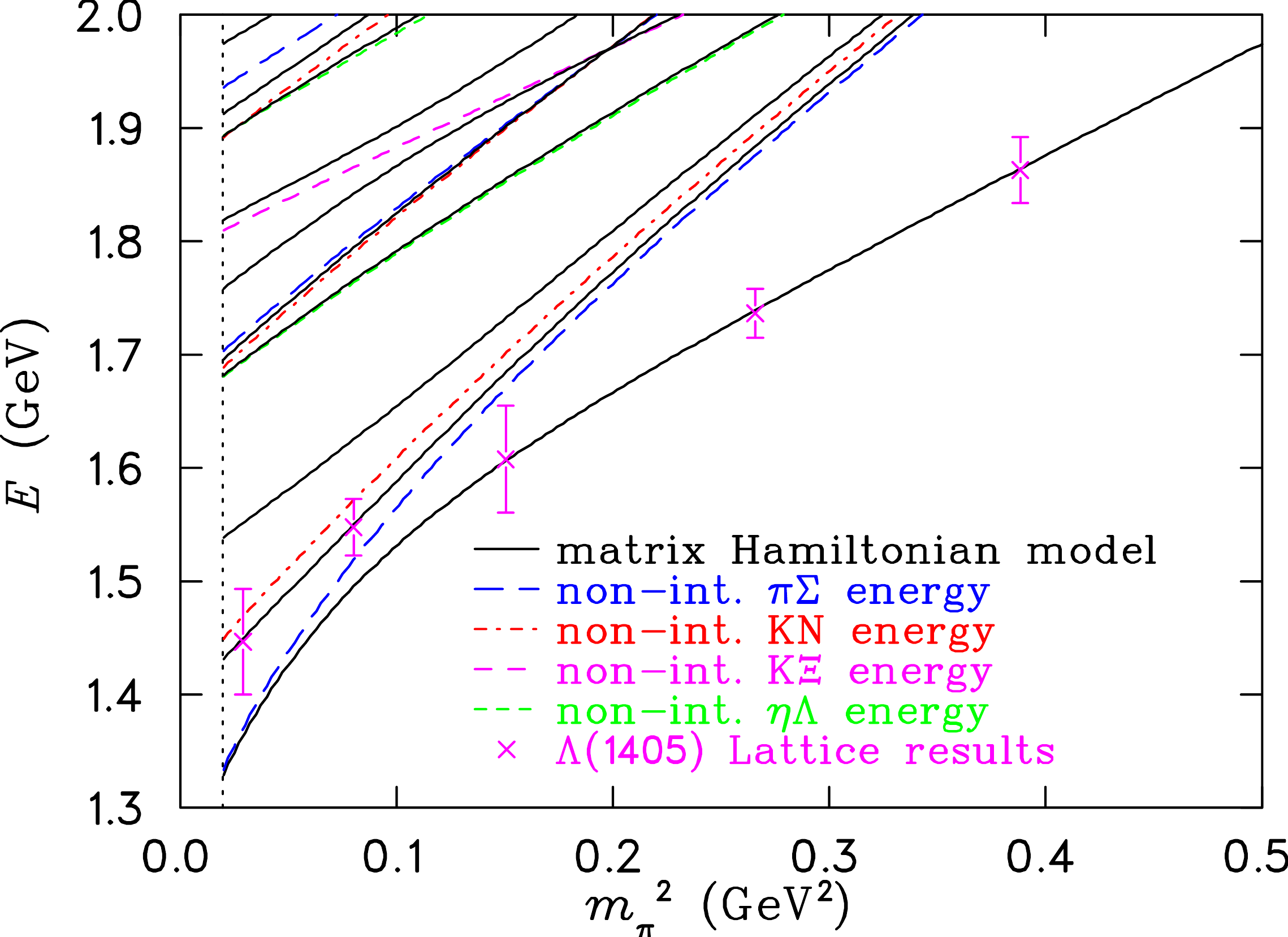}
\quad
\includegraphics[height=0.48\columnwidth,angle=90]{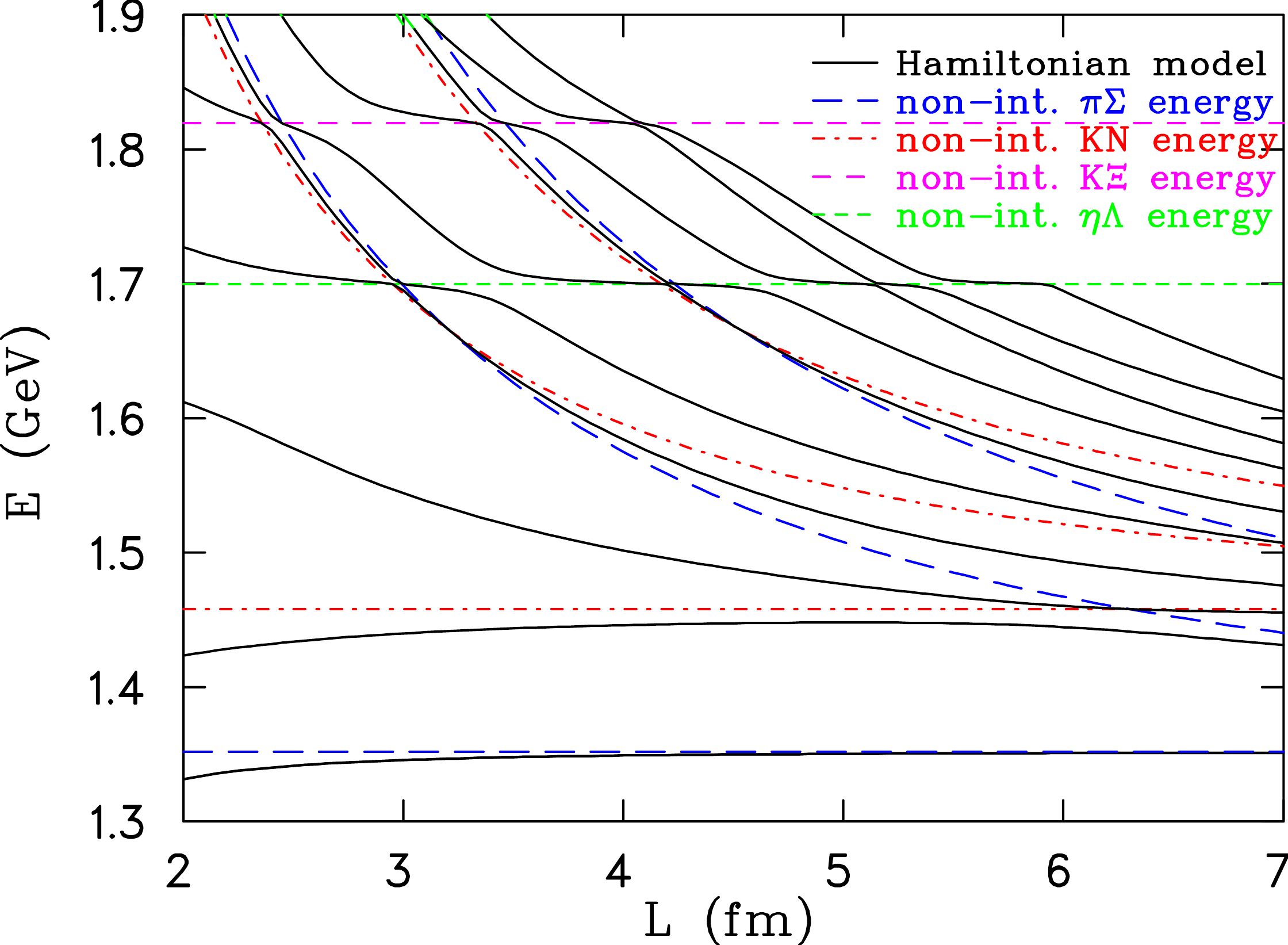}
\vspace*{-8pt}
\caption{{\bf (left)} The quark-mass dependence $(m_q \propto
  m_\pi^2)$ of the lowest-lying $\Lambda(1405)$ states observed in our
  lattice QCD calculations \cite{Hall:2014lambda} is illustrated by
  the discrete points at each of the pion masses available in the
  PACS-CS ensembles.  The low-lying energy spectrum of our Hamiltonian
  model (solid curves) constrained to the Lattice QCD results
  (discrete points) is also illustrated.  The associated
  non-interacting meson-baryon basis states are illustrated by the
  dashed curves.
{\bf (right)} The volume dependence of the spin-1/2 odd-parity $\Lambda$
  spectrum obtained from our Hamiltonian effective field theory
  analysis of our lattice results at $m_\pi = 156$ MeV.  $L$ denotes
  the length of the $L^3$ volume.
\label{fig:modelfin}}
\end{figure}

To solve this Hamiltonian system, the {\em dgeev} routine from the LAPACK
software library is used to obtain the eigenvalues and eigenvectors of
$H$. The energy eigenvalues of the matrix may be fitted to their
corresponding lattice QCD values by minimising the chi-square function
for the parameters $m_0$ and $\alpha_0$ at different
values of $m_\pi^2$.  The low-lying energy eigenvalues fit to the
lattice QCD results are illustrated in Fig.~\ref{fig:modelfin}.
The scale is set via the Sommer parameter 
with $r_0 = 0.492$ fm \cite{Aoki:2008sm}.

\section{Discussion}

%
The three heaviest quark masses considered on the lattice correspond
to a stable odd-parity $\Lambda(1405)$, as the $\pi\Sigma$ threshold
energy exceeds that of the $\Lambda(1405)$.  However, as the physical
pion mass is approached, the $\pi\Sigma$ threshold energy decreases
and a nontrivial mixing of states associated with an avoided level
crossing of the transitioning $\pi\Sigma$ threshold occurs.  At the
lightest two quark masses considered, the $\Lambda(1405)$ corresponds
to the second state of the Hamiltonian model with a
$\pi\Sigma$-dominated eigenstate occupying the lowest energy position.

%
The right-hand panel of Fig.~\ref{fig:modelfin} presents the volume
dependence of the odd-parity $\Lambda$ spectrum obtained from our
Hamiltonian effective field theory analysis of our lattice results at
the lightest quark mass providing $m_\pi = 156$ MeV.  The second
state, identified as the $\Lambda(1405)$ via the quark mass dependence
of the spectrum presented in the left-hand panel of
Fig.~\ref{fig:modelfin}, undergoes an avoided level crossing just
beyond $L = 6$ fm to become the third state in the spectrum.  Here the
second state turns down to approach the $\pi\Sigma$ scattering state
where each hadron carries one unit of lattice momenta, $2\pi/L$.  The
process of avoided level crossings will repeat as more $\pi\Sigma$
scattering states have their energy drop below the $\Lambda(1405)$
energy as $L$ increases.

%
The correlation matrix approach~\cite{Michael:1985ne,Luscher:1990ck}
was used to determine the superposition of interpolators required to
isolate the $\Lambda(1405)$ on the lattice \cite{Menadue:2011pd}.
This discovery of a low-lying $\Lambda(1405)$ mass
\cite{Menadue:2011pd} has since been independently confirmed
\cite{Engel:2012qp}.
The success of our approach in accurately isolating the
$\Lambda(1405)$, even at the lightest quark mass considered, is best
demonstrated by the long Euclidean-time single-state stability of the
$\Lambda(1405)$ two-point correlation function presented in Fig.~2 of
Ref.~\cite{Menadue:2013xqa}.  We note this success is realised without
resort to two-particle interpolators.  

%
The results of Fig.~\ref{fig:modelfin} emphasize how the near-by
$\overline{K}N$ threshold realised in the infinite-volume limit of
Nature does not sit next to the $\Lambda(1405)$ in a finite volume
with $L \sim 3$ fm.  Rather the mixing of states in the finite volume
renders any infinite-volume concerns on the role of nearby thresholds
irrelevant.

%
The vanishing of the strange quark contribution to the magnetic form
factor of the state observed in our lattice calculation at the
lightest quark mass signals the creation of a $\overline{K}N$ state
from three-quark operators.  The success of creating a five-quark
state from a three quark operator is reminiscent of the dream of
observing string breaking in the static quark potential.  There the
hope was that the sea-quark loops in the vacuum would become manifest
in a flattening of the potential at large separations.  However the
dream was not realised and explicit quark degrees of freedom had to be
introduced by hand in a correlation matrix analysis.
In contrast, the sea-quark loops in the vacuum are manifest in the
magnetic form factor of the $\Lambda(1405)$, {\em without} resort to
explicit five-quark operators.

\section{Infinite-Volume Limit}

%
Having confirmed that the $\Lambda(1405)$ state observed on the
lattice is best described as a molecular $\overline{K}N$ bound state,
it remains to demonstrate the connection between the finite-volume
lattice eigenstates and the infinite-volume resonance found in Nature.
The quark-mass behaviour of the $\Lambda(1405)$ energy in the
infinite-volume limit can be reconstructed from the finite-volume
Hamiltonian model by considering the principal-value continuum
versions of the loop integral contributions from all channels with the
appropriate physical hadron masses.  The resonance energy of the
$\Lambda(1405)$ in infinite volume is
\begin{equation}
E_{\Lambda1405} 
=m_0^{\mathrm{fit}} + \alpha_0^{\mathrm{fit}}\, m_\pi^2 
+\sum_{M,B}
\frac{\kappa_{MB}}{16\pi^2 f_\pi^2}\mathcal{P}\!\int_0^\infty\!
\mathrm{d}k\,
\frac{k^2\, \omega_M(k)\, u^2(k)}{E_{\Lambda1405} - \omega_{MB}(k)}, 
\label{eq:infvolcalc}
\end{equation}
where $\mathcal{P}$ indicates that the principal value integral is
performed.  This integral represents the infinite-volume version of the
loop sum appearing in Eq.~(\ref{eq:eval1}).  Since $E_{\Lambda1405}$
appears on both sides of Eq.~(\ref{eq:infvolcalc}), it is best solved
iteratively by scanning over a range of possible values of
$E_{\Lambda1405}$.

\begin{figure}[t]
\begin{minipage}{\textwidth}%
\begin{minipage}{0.5\textwidth}
\centering
\includegraphics[height=0.98\columnwidth,angle=90]{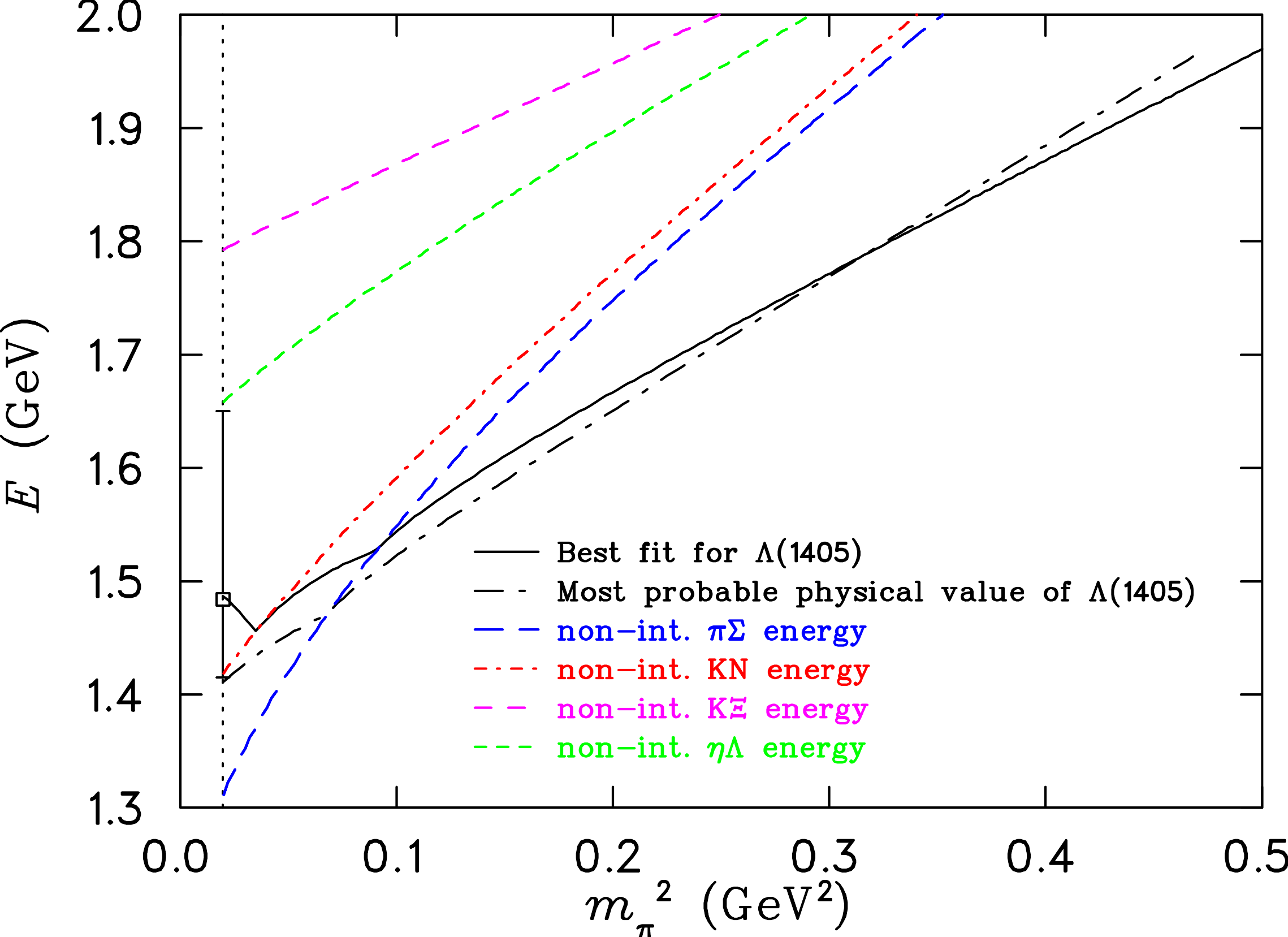}
\end{minipage}%
\ 
\begin{minipage}{0.5\textwidth}
\includegraphics[width=0.98\columnwidth,angle=0]{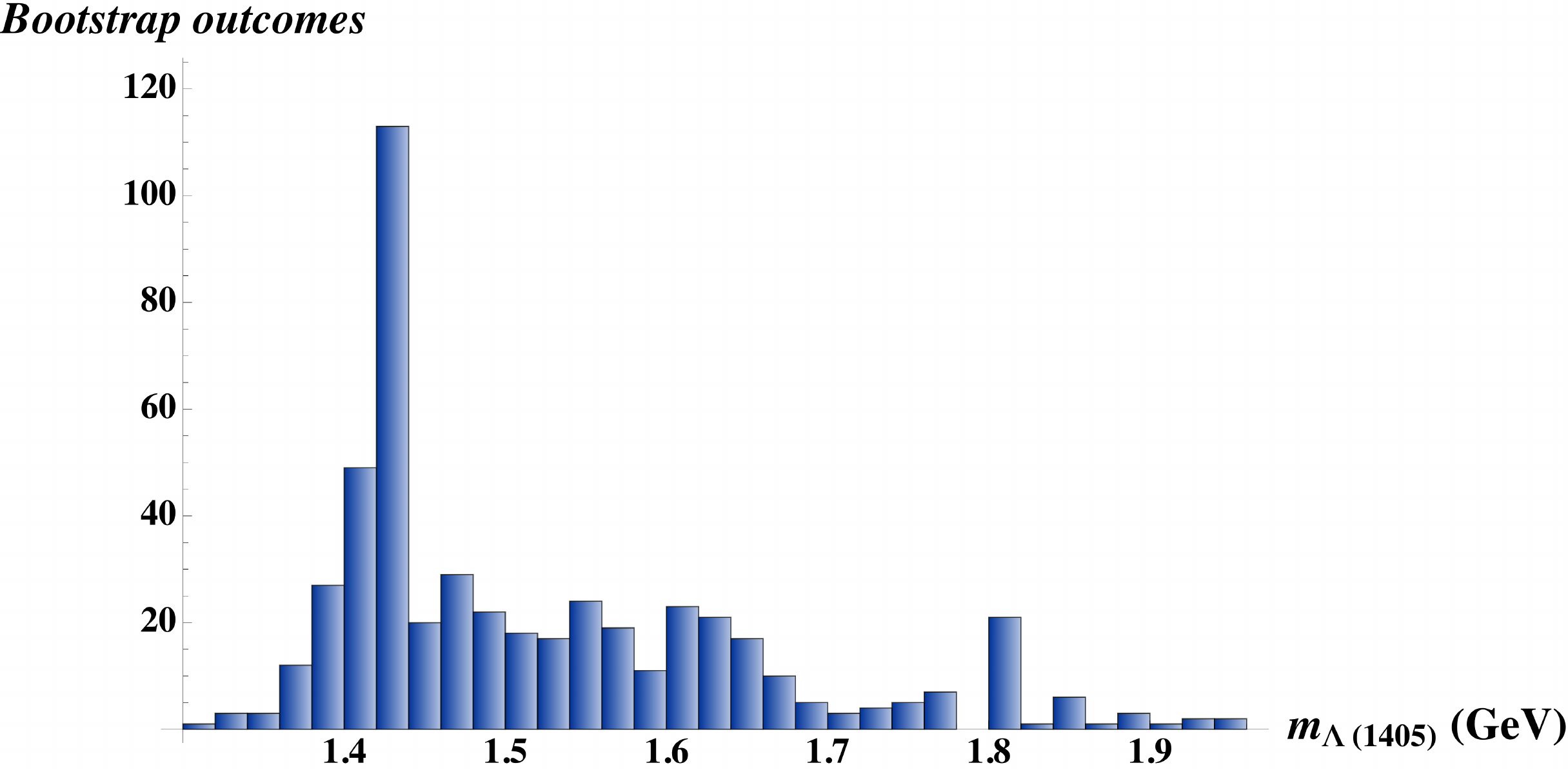}
\end{minipage}%
\end{minipage}
\caption{{\bf (left)} The quark mass dependence of the infinite-volume
  reconstruction of the $\Lambda(1405)$ energy is
  illustrated as a function of the squared pion-mass, $m_\pi^2$.  Both
  the fit to the central values of the lattice QCD results and the
  most probable mass dependence observed in the bootstrap ensemble
  analysis of the lattice results are illustrated.
  {\bf (right)} The statistical distribution of the infinite-volume
  $\Lambda(1405)$ resonance energy at the physical pion mass obtained
  from a bootstrap analysis of the Lattice QCD results.
\label{fig:modelinf}}
\end{figure}

The result is shown in the left-hand panel of Fig.~\ref{fig:modelinf}
as a solid black line.  The dashed lines illustrate the
non-interacting infinite-volume $S$-wave threshold energies, which
each induce a nonanalytic cusp in the quark-mass dependence of the
self energy.

%

To obtain an estimate of the statistical uncertainty in the
$\Lambda(1405)$ energy, a bootstrap analysis is performed.  This is
achieved by repeating the minimisation of the chi-square to obtain
fitted values of $m_0$ and $\alpha_0$ for separate bootstrap ensembles
of lattice QCD data.  The bootstraps are calculated by altering the
value of each lattice data point by a Gaussian-distributed random
number, weighted by the uncertainty at each point in $m_\pi^2$.
The statistical distribution of values for the $\Lambda(1405)$ mass at
the physical pion mass and infinite volume, for 500 bootstrap
configurations, is shown in the form of a histogram, displayed in the
right-hand panel of Fig.~\ref{fig:modelinf}.  The plot has an
unconventional distribution due to cusps in the extrapolation
associated with the opening of decay channel(s).  The bootstrap error
analysis provides a resonance energy of $1.48{+0.17 \atop -0.07}$ GeV.

The distribution of the bootstrap analysis is sharply peaked around
the most probable value of 1.41 GeV in good agreement with experiment.
We illustrate the most probable outcome for the $\Lambda(1405)$ mass
dependence by the dot-dash curve in the left-hand panel of
Fig.~\ref{fig:modelinf} labeled ``Most probable physical value of
$\Lambda(1405)$.''  In this case only the $\pi\Sigma$ threshold
induces a cusp and the ordering of the $\pi\Sigma$ threshold, the
$\Lambda(1405)$ energy, and the $\overline{K}N$ threshold realised in
Nature is reproduced.

\section{Wave Function Renormalisation}

%
The pion-mass dependence of the $\Lambda(1405)$ can be further
examined via the wave function renormalisation, $Z_2$, defined as the
ratio of the renormalised coupling constant to the bare coupling
constant.  It can be calculated using\hfill\null
\begin{wrapfigure}{r}{0.48\hsize}
\vspace*{-12pt}
\begin{center}
\includegraphics[height=0.48\columnwidth,angle=90]{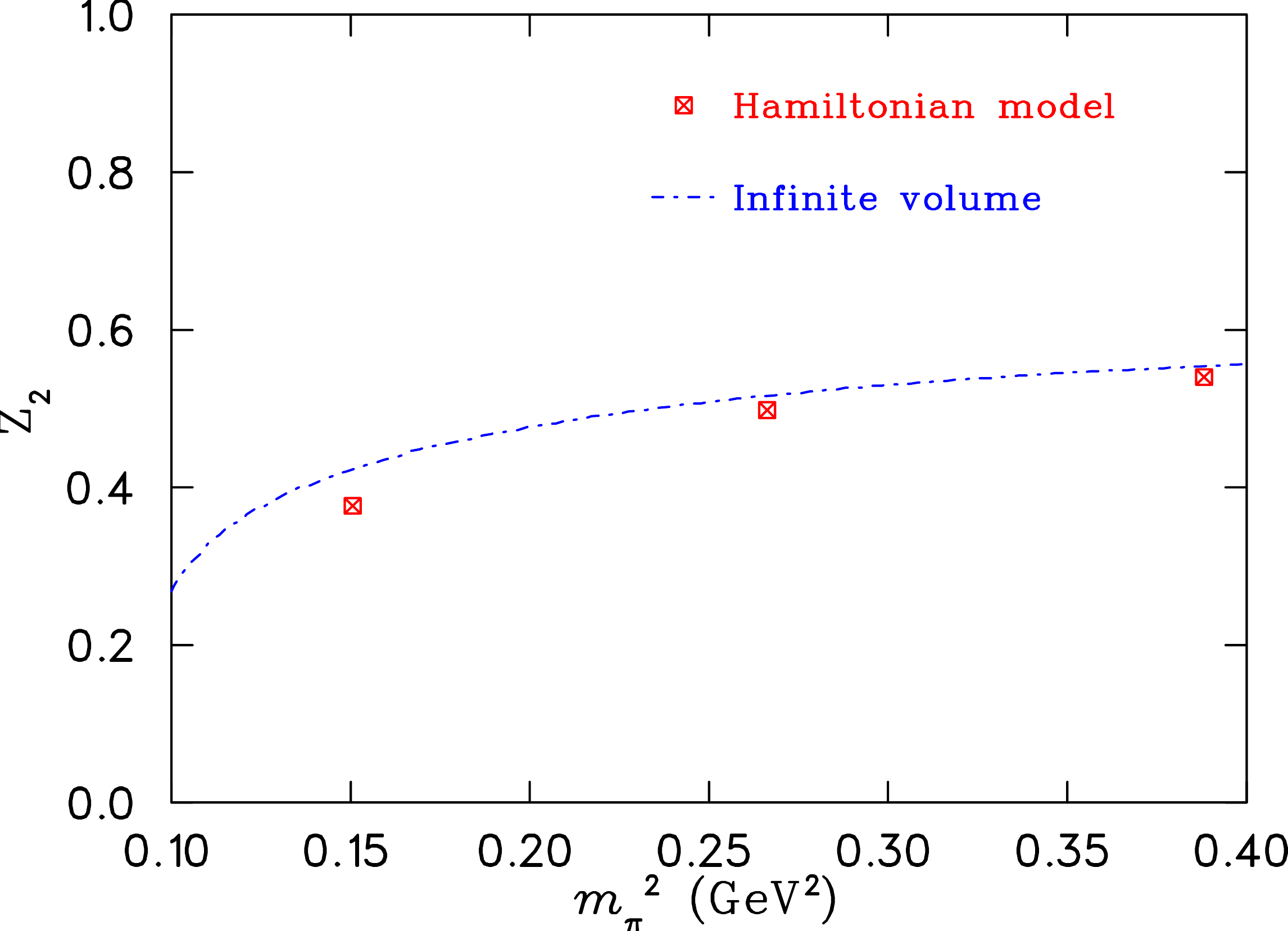}
\caption{{Wave function renormalisation factor
    \boldmath{$Z_2$} as a function of pion mass.}  While $Z_2=1$
  corresponds to an elementary particle, $Z_2 = 0$ corresponds to a
  bound state. For $m_\pi < 216$ MeV, the infinite-volume $Z_2$ cannot
  be defined.
\label{fig:Z2}}
\end{center}
\vspace*{-18pt}
\end{wrapfigure}
\begin{equation}
Z_2 = \frac{1}{1-\frac{\partial}{\partial E}\Sigma^{\mathrm{loops}}(m_\pi^2,E)} 
\Big|_{E = E_{\Lambda1405}}. 
\end{equation}
This quantity describes the extent to which the resonance behaves like
an elementary particle where $Z_2 = 1$, or a bound state where $Z_2 =
0$.  

Results for $Z_2$ are illustrated in Fig.~\ref{fig:Z2}.  
For large pion masses, the $\Lambda(1405)$ is more like an elementary
particle.  In addition, the finite- and infinite-volume values of
$Z_2$ match very closely. This is to be expected in a quark-mass
region that lies far from a resonance.  As the pion mass becomes
smaller, the Hamiltonian model suggests stronger overlap with
neighbouring particle states associated with a molecular bound state.
The probability of the 3-quark component decreases and the eigenstate
is predominantly a $\bar{K}N$ bound state.  For $m_\pi < 216$ MeV, the
$\pi\Sigma$ channel opens, and the infinite-volume wave-function
renormalisation can no longer be defined in the resonant region.


\begin{thebibliography}{10}

\bibitem{Hall:2014lambda}
J.~M.~M. Hall, W.~Kamleh, D.~B. Leinweber, B.~J. Menadue, B.~J. Owen, A.~W.
  Thomas, and R.~D. Young, \null \href{http://arxiv.org/abs/1411.3402}{{\tt
  arXiv:1411.3402}}.

\bibitem{Engler:1965zz}
A.~Engler, H.~Fisk, R.~Kraemer, C.~Meltzer, and J.~Westgard, \null {\em Phys.\
  Rev.\ Lett.} {\bf 15} (1965) 224.

\bibitem{Hall:2013qba}
J.~Hall, A.~C.~P. Hsu, D.~Leinweber, A.~Thomas, and R.~Young, \null {\em Phys.\
  Rev.} {\bf D87} (2013) 094510, [\href{http://arxiv.org/abs/1303.4157}{{\tt
  arXiv:1303.4157}}].

\bibitem{Leinweber:2003dg}
D.~B. Leinweber, A.~W. Thomas, and R.~D. Young, \null {\em Phys.\ Rev.\ Lett.}
  {\bf 92} (2004) 242002, [\href{http://arxiv.org/abs/hep-lat/0302020}{{\tt
  hep-lat/0302020}}].

\bibitem{Veit:1984an}
E.~Veit, B.~K. Jennings, R.~Barrett, and A.~W. Thomas, \null {\em Phys.\ Lett.}
  {\bf B137} (1984) 415.

\bibitem{Veit:1984jr}
E.~Veit, B.~K. Jennings, A.~W. Thomas, and R.~Barrett, \null {\em Phys.\ Rev.}
  {\bf D31} (1985) 1033.

\bibitem{Beringer:1900zz}
{\bf Particle Data Group} Collaboration, J.~Beringer et~al., \null {\em Phys.\
  Rev.} {\bf D86} (2012) 010001.

\bibitem{Aoki:2008sm}
{\bf PACS-CS Collaboration} Collaboration, S.~Aoki et~al., \null {\em Phys.\
  Rev.} {\bf D79} (2009) 034503, [\href{http://arxiv.org/abs/0807.1661}{{\tt
  arXiv:0807.1661}}].

\bibitem{Michael:1985ne}
C.~Michael, \null {\em Nucl.\ Phys.} {\bf B259} (1985) 58.

\bibitem{Luscher:1990ck}
M.~Luscher and U.~Wolff, \null {\em Nucl.\ Phys.} {\bf B339} (1990) 222--252.

\bibitem{Menadue:2011pd}
B.~J. Menadue, W.~Kamleh, D.~B. Leinweber, and M.~S. Mahbub, \null {\em Phys.\
  Rev.\ Lett.} {\bf 108} (2012) 112001,
  [\href{http://arxiv.org/abs/1109.6716}{{\tt arXiv:1109.6716}}].

\bibitem{Engel:2012qp}
G.~P. Engel, C.~Lang, and A.~Schafer, \null {\em Phys.\ Rev.} {\bf D87} (2013),
  no.~3 034502, [\href{http://arxiv.org/abs/1212.2032}{{\tt arXiv:1212.2032}}].

\bibitem{Menadue:2013xqa}
B.~J. Menadue, W.~Kamleh, D.~B. Leinweber, M.~S. Mahbub, and B.~J. Owen, \null
  {\em PoS} {\bf LATTICE2013} (2013) 280,
  [\href{http://arxiv.org/abs/1311.5026}{{\tt arXiv:1311.5026}}].

\end{thebibliography}

\providecommand{\href}[2]{#2}\begingroup\raggedright\endgroup

\end{document}